\documentclass[10pt,aps,prd,nofootinbib,superscriptaddress,twocolumn,preprintnumbers,balancelastpage]{revtex4}
\usepackage{amssymb,amsmath,latexsym,graphics, graphicx,epsfig,multirow,comment,hyperref,appendix}

\newcommand {\be} {\begin {equation}}
\newcommand {\ee} {\end {equation}}
\newcommand {\ba} {\begin {eqnarray}}
\newcommand {\ea} {\end {eqnarray}}

\newcommand {\bes} {\begin {equation*}}
\newcommand {\ees} {\end {equation*}}

\newcommand{\es}[2] {\begin{equation} \label{#1} \begin{split} #2 \end{split} \end{equation}}

\def\lsim{\mathrel {\vcenter {\baselineskip 0pt \kern 0pt
    \hbox{$<$} \kern 0pt \hbox{$\sim$} }}}
\def\gsim{\mathrel {\vcenter {\baselineskip 0pt \kern 0pt
    \hbox{$>$} \kern 0pt \hbox{$\sim$} }}}

\newcommand{\beq}{\begin{equation}}
\newcommand{\eeq}{\end{equation}}

\begin{document}
\title{Constraining Axion Dark Matter with Big Bang Nucleosynthesis}
\author{Kfir Blum}
\affiliation{Institute for Advanced Study, Princeton, NJ 08540}
\author{Raffaele Tito D'Agnolo}
\affiliation{Institute for Advanced Study, Princeton, NJ 08540}
\author{Mariangela Lisanti}
\affiliation{Department of Physics, Princeton University, Princeton, NJ 08544}
\author{Benjamin R. Safdi}
\affiliation{Department of Physics, Princeton University, Princeton, NJ 08544}

\date{\today}

\begin{abstract}{
We show that Big Bang Nucleosynthesis (BBN) significantly constrains axion-like dark matter.
The axion acts like an oscillating QCD $\theta$ angle that redshifts in the early universe, increasing the neutron-proton mass difference at neutron freeze-out. An axion-like particle that couples too strongly to QCD results in the underproduction of $^4$He during BBN and is thus excluded.  The BBN bound overlaps with much of the parameter space that would be covered by proposed searches for a time-varying neutron EDM.  The QCD axion does not couple strongly enough to affect BBN.  
}

\end{abstract}
\maketitle

The axion is a well-motivated dark-mater (DM) candidate that can arise in a variety of models~\cite{Beringer:1900zz}.  The allowed mass of these light scalars is relatively unconstrained, spanning many orders of magnitude.  Identifying the regions of axion parameter space that are excluded by cosmological and astrophysical constraints is of the utmost importance as it directs the focus of laboratory searches.  This Letter presents a new constraint on axion dark matter arising from Big Bang Nucleosynthesis. 

The axion was originally introduced to explain why the QCD $\theta$ term,
\es{tQCD}{
S=\frac{\theta}{4\pi^2}\int{\rm tr}~G\wedge G \, ,
} 
is not realized in Nature, often referred to as the strong CP problem~\cite{PhysRevLett.38.1440,Weinberg:1977ma,Wilczek:1977pj}.\footnote{Our conventions are $\int{\rm tr}~G\wedge G=(1/4)\int d^4x\epsilon^{\mu\nu\alpha\beta}{\rm tr}~G_{\mu\nu}G_{\alpha\beta}$, where $G=(1/2)G_{\mu\nu}dx^\mu\wedge dx^\nu$ is the gluon field-strength with the trace taken over gauge indices. In the following, we use also $\tilde G^{\mu\nu}=\epsilon^{\mu\nu\alpha\beta}G_{\alpha\beta}$.}  The QCD $\theta$ term in~\eqref{tQCD} induces a neutron electric dipole moment (EDM) \mbox{$d_n \approx 2.4 \times 10^{-16}$ $\theta$ e\;cm}~\cite{Pospelov:1999ha} that is in tension with experiment for $\theta>10^{-10}$~\cite{Baker:2006ts,Harris:1999jx}. The axion solves this problem by promoting the parameter $\theta$ to a dynamical field, $\theta\to (a/f_a)$, whose potential is minimized at $a = 0$. 

The axion is often assumed to be the pseudo-Goldstone boson of a $U(1)$ PQ symmetry, which is spontaneously broken at some high scale, $f_a$~\cite{Weinberg:1977ma,Wilczek:1977pj,Kim:1979if,Dine:1981rt}.  For the axion to solve the strong CP problem, the explicit breaking of the PQ symmetry must be absent to very high accuracy in the UV~\cite{Holman:1992us,Kallosh:1995hi}. The leading potential that the axion is allowed to receive should come from the QCD chiral anomaly. The QCD instantons break the PQ symmetry explicitly, and in the presence of bare quark masses, the axion picks up a mass~\cite{Weinberg:1977ma,Wilczek:1977pj}
\es{massQCD}{
f_ a \,m_a =f_\pi m_\pi { \sqrt{m_u m_d } \over m_u + m_d}  \,,
} 
where $m_\pi \approx 140$ MeV is the pion mass, $f_\pi \approx 92$ MeV is the decay constant, and $m_u \approx$ 2.3 MeV ($m_d \approx 4.8$ MeV) is the mass of the up (down) quark.

The cosmological equation of state of axion DM is governed by the classical oscillations of the background field~\cite{Abbott:1982af,Preskill:1982cy,Dine:1982ah,Turner:1985si}: 
\es{eqState}{
a(t) = a_0 \cos (m_a t) = \frac{\sqrt{2 \rho_\text{DM}}}{m_a} \cos(m_a t) \,\, .
}
The amplitude $a_0$ is fixed by requiring that the axion makes up the observed dark matter density, $\rho_\text{DM}$.  A parameter space spanning orders of magnitude in $m_a$ and $f_a$ is available to axion DM.  
Constraints on axions that come from their coupling to $G \wedge G$ arise from excess cooling of SN 1987A~\cite{Graham:2013gfa,Raffelt:2006cw} 
and from static neutron EDM measurements~\cite{Baker:2006ts,Harris:1999jx,Graham:2013gfa}.  Axions may also be constrained through their coupling to $E \cdot B$ (see~\cite{Beringer:1900zz} for a review).

In addition to the QCD axion, axion-like particles (ALPs) can arise in many models. ALPs do not necessarily couple to $G \wedge G$; for example, they may only couple to electromagnetism through the operator $E \cdot B$. In these models,~\eqref{massQCD} may be violated, and in particular, it is possible that 
\es{firstSpace}{
f_a m_a \ll \Lambda_\text{QCD}^2 \,.
} 
From this point forward, we will use ``axion" to refer to both the standard QCD axion and ALPs that couple to $G \wedge G$ with coupling $\propto f_a^{-1}$ and that satisfy~\eqref{firstSpace}.

Axions that couple to $G \wedge G$ and simultaneously satisfy~\eqref{firstSpace} may be tested directly in the near future by proposed laboratory searches for an oscillating axion-induced nucleon EDM~\cite{Graham:2011qk,Budker:2013hfa,Graham:2013gfa}.  This Letter focuses on this region of axion parameter space.  First, we review a well-known result from chiral perturbation theory (ChPT), showing that the presence of an axion-induced nucleon EDM is in tension with~\eqref{firstSpace} because the axion contribution to the nucleon EDM is associated with the irreducible QCD contribution to the axion mass in~\eqref{massQCD}.  As far as we know, the only way to avoid this minimum axion mass is to invoke fine-tuned cancellations, exacerbating the strong CP problem. 
 
Even if one is willing to ignore fine-tuning arguments, Big Bang Nucleosynthesis (BBN) provides a strong observational constraint.  The constraint arises from two simple observations.  First, the QCD $\theta$ term leads to a shift in the neutron-proton mass difference, as pointed out in~\cite{Ubaldi:2008nf}.  This nuclear mass difference is again dictated by ChPT and is directly related to the axion-induced EDM. Second, the effective $\theta$ term induced by axion DM redshifts in the early universe, roughly as $\theta\sim(1+z)^{3/2}$.  Thus, while the effect of axion DM on the neutron-proton mass difference today seems unobservably small, it can be large enough to disturb the production of light elements at the time of BBN (redshift $z\sim10^{10}$).   

We begin by recalling the results from ChPT that relate the axion mass and some of its couplings. 
Considering only the axion and strongly-interacting SM fields just above the QCD scale, the most general effective Lagrangian that connects the axion to the SM and respects the axion shift symmetry is
\be\label{eq:Leff0} \mathcal{L}=-\frac{a}{f_a}\frac{G_{\mu\nu}^a \tilde G^{a \mu\nu}}{32\pi^2}-\frac{\partial_\mu a}{f_a}\sum_{\psi}c_\psi\bar \psi\bar\sigma^\mu \psi\ee
to leading order in $f_a^{-1}$~\cite{Georgi:1986df}.  The left-handed Weyl spinors $\psi$ include $u,u^c,d,d^c$ etc. They are allowed to have derivative couplings 
to the axion $a$, with model-dependent coefficients $c_\psi$ that may be off-diagonal in flavor space.
Below the QCD scale, (\ref{eq:Leff0}) is translated to the chiral Lagrangian. This is done most easily by first performing a spacetime-dependent chiral phase redefinition of the quark fields to eliminate the $G \tilde G$ term and replacing it by a complex phase in the quark mass matrix along with finite shifts in the coefficients $c_\psi$.

From this point on, the axion couplings with pions and nucleons may be computed from ordinary ChPT.  The axion enters into the chiral Lagrangian only through the quark mass spurion and through mixed derivative couplings with the neutral pion. Working in the physical basis after diagonalizing the axion-pion mass matrix and kinetic terms, 
 we are particularly interested in the following terms in the chiral Lagrangian~\cite{Crewther:1979pi,Ubaldi:2008nf}:
\es{eq:Lch}{
\mathcal{L}\supset& - {1 \over 2} {f_\pi^2 m_\pi^2 m_u m_d \over (m_u + m_d)^2} \left({ a \over f_a }\right)^2 \\
-&\bar N\pi\cdot\sigma \left(i \gamma^5 g_{\pi NN} - 2 \, \bar g_{ \pi NN}\frac{a}{f_a}\right)N \\
+&\frac{f_\pi \, \bar g_{ \pi NN}}{2}\frac{m_d-m_u}{m_d+m_u}\left(\frac{a}{f_a}\right)^2\,\bar N\sigma^3 N \,.
}
Here
$N=\left(\begin{array}{c}p\\n\end{array}\right)$
 are the nucleons, and
the 
numerical 
couplings are $g_{\pi NN}\approx13.5$ and $\bar g_{\pi NN}\approx\frac{m_um_d}{m_u+m_d}\frac{2(M_\Xi-M_\Sigma)}{(2 m_s - m_u - m_d)f_\pi}\approx0.023$.  

The first line in~(\ref{eq:Lch}) is the irreducible contribution to the axion mass quoted in~(\ref{massQCD}). We know of no way to eliminate this contribution for an axion with decay constant $f_a$ besides to cancel it with some unrelated mass correction associated with some new Lagrangian term $\Delta\mathcal{L}(a) \propto \delta m^2 (a + \delta \theta)^2$. Such a cancelation would involve fine-tuning the parameter $\delta m^2$ by an amount
\es{tuning}{
\Delta_{\text mass} \sim\frac{f_a^2\,m_a^2}{f_\pi^2 m_\pi^2}\sim10^{-14}\,\left(\frac{f_a\,m_a}{10^{-9}~{\rm GeV^2}}\right)^{2} \,.
}
Moreover, $\delta \theta$ must also be tuned to not spoil the solution to the strong CP problem,
  thereby restoring it on top of the mass fine-tuning in~\eqref{tuning}. 

The second line in~(\ref{eq:Lch}) gives the dominant contribution to the  axion-induced neutron EDM~\cite{Crewther:1979pi,Graham:2011qk},
\be d_n \approx \left( \frac{a}{f_a} \right) \frac{eg_{\pi NN}{\bar g}_{\pi NN}}{4\pi^2}\frac{\ln(4 \pi f_\pi /m_\pi)}{m_N} \,,\ee
with $m_N$ the nucleon mass in the limit of vanishing up and down quark masses.

The third line in~(\ref{eq:Lch}) gives the axion-induced neutron-proton mass splitting,
\es{mnmpDiff}{
&m_n - m_p = Q_0 + \delta Q  \,, \\
&\delta Q \approx { f_\pi \, {\bar g}_{\pi NN} \over 2} \left( { m_d - m_u \over m_d + m_u} \right) \left( {a \over f_a} \right)^2 \\
&\approx  \big( 0.37 \, \, \text{MeV} \big) \left( {a \over f_a} \right)^2 \,,
}
when evaluated on a classical axion-field background.
$Q_0 \approx 1.293$ MeV is the measured mass difference between the neutron and proton.
Thus, an axion field that induces a nuclear EDM also affects the neutron-proton mass splitting in a directly related way.
Moreover, the relation between the two effects does not depend on the model-dependent $c_\psi$ coefficients, to leading order in $1/f_a$.\footnote{The relation between the nuclear EDM and the neutron-proton mass splitting could be modified if we allow for other sources of explicit PQ symmetry breaking beyond the mass-tuning term.  We do not consider such possibilities in this Letter.}   
We now explore the consequence of the shift in the nuclear mass difference on nucleosynthesis.

For $m_a \gg H(z)$, where $H(z)$ is the proper Hubble expansion rate at redshift $z$, the axion DM may be treated as an ensemble of Bose-Einstein condensed non-relativistic particles~\cite{Turner:1985si}.  Neglecting any temperature dependence in $m_a$, the time-dependent effective $\theta$ angle in this limit is   
\es{thetaEff}{
&\theta_\text{eff}(t) = (1+z(t))^{3/2} { \sqrt{2 \bar \rho_\text{DM}} \over f_a \, m_a}  \cos (m_a t)  \\
&\approx 5 \times 10^{-9} \left( {\text{GeV}^2 \over f_a m_a } \right) \left( {1+z(t) \over 10^{10}} \right)^{3/2} \cos(m_a t) \,,
}     
where $ \bar \rho_\text{DM} \approx 2.7 \times 10^{-27}$ kg$/$m$^3$ is the mean cosmological DM energy density today~\cite{Ade:2013zuv}.  
Neutron freeze-out occurs at temperatures of order 1~MeV, meaning that~\eqref{thetaEff} is adequate for calculating a BBN bound as long as $m_a \gg  \big(1 \, \, \text{MeV} \big)^2 / m_\text{pl} \approx 10^{-16} \, \, \text{eV}$. We begin by discussing $m_a$ in this regime and extend the calculation to the ultra-light regime, $m_a\ll10^{-16}$~eV, later.  

Substituting~\eqref{thetaEff} into~\eqref{mnmpDiff} shows that axion DM increases the mass difference between the neutron and proton at BBN.  This reduces the relative occupation number of neutrons compared to that of protons in thermal equilibrium just before neutron freeze-out, reducing the resulting mass fraction, $Y_p$, of $^4$He. The net effect is stronger at smaller $f_a \, m_a$.  We now provide an analytic estimate of the dependence of $Y_p$ on $f_a \, m_a$, subsequently moving on to a more precise numerical calculation. 

After the quark-hadron transition, neutrons and protons are kept in equilibrium through the weak interactions 
\es{weakR}{
n \, &\longleftrightarrow  \,p + e^{-} + \bar \nu_e \,, \\
\nu_e + n \, &\longleftrightarrow  \,p + e^{-}  \,, \\
e^+ + n \, &\longleftrightarrow  \,p +  \bar \nu_e \,.
}
The rates of these reactions become smaller than the Hubble parameter around the freeze-out temperature $T_\text{F} \approx 0.8$ MeV.  Below this temperature, neutrons and protons fall out of equilibrium, and the neutron to proton ratio is approximately fixed to the ratio of $n/p$ at freeze-out:
\es{npE}{
\left( {n \over p} \right)_\text{freeze-out} \approx e^{-Q_\text{F} / T_\text{F}} \,,
}
where $Q_F = Q_0 + \delta Q_\text{F}$ is the mass difference between the neutron and proton at freeze-out. 

For $m_a > 10^{-16}$ eV, the axion oscillation frequency $m_a$ is greater than the rate of the weak interactions in~\eqref{weakR} when $T \approx T_\text{F}$.  
Each weak scattering event therefore sees a different value for $Q_{\rm F}$, and $Q_\text{F}$ 
in~\eqref{npE} should be averaged over times of order $m_a^{-1}$. This amounts to replacing the factor $\cos^2(m_a t)$ by a $1/2$ when using~\eqref{thetaEff}.\footnote{We assume that the nucleon distribution functions just before neutron freeze-out achieve a quasi-equilibrium state, dictated by the respective nucleon mass averaged over a time interval of order $m_a^{-1}$.  } 

In addition to the direct effect on $Q_{\rm F}$, decreasing $f_a \, m_a$ also decreases the freeze-out temperature $T_F$ itself.  This effect is small in the range of $f_a \, m_a$ of interest here.  For now,  we assume that $T_F$ $\sim 0.8$ MeV is unchanged and relax this assumption later in the numerical calculation.  

All of the neutrons left over at the end of the deuterium bottleneck, which occurs at a temperature $T_\text{Nuc} \approx 0.086$ MeV, are converted into $^4$He, to a good approximation.  The mass fraction of $^4$He is approximately
\es{Yp}{
Y_p \approx {2 (n/p)_\text{Nuc} \over 1 + (n/p)_\text{Nuc} } \,.
}
Between freeze-out and nucleosynthesis, a small fraction of neutrons are lost by free decay.  To a first approximation, we neglect neutron decay and estimate the fractional change in $Y_p$ as a result of the axion DM,
\es{deltaYp}{
{\delta Y_p \over Y_p} \equiv {Y_p^0 - Y_p(f_a \, m_a ) \over Y_p^0}  \,,
}
by taking $(n/p)_\text{Nuc} \approx (n/p)_\text{freeze-out}$ and using~\eqref{npE}.   $Y_p^0$ denotes the value of $Y_p$ in the absence of axion DM.   

The mass fraction $Y_p^0$ of $^4$He is measured to be in the range 0.227--0.266~\cite{Beringer:1900zz}, to 95\% confidence.
Taking the conservative bound $\delta Y_p / Y_p < 10$\%, we find the constraint $f_a \, m_a \gsim  10^{-9} $ GeV$^2$.
 \begin{figure*}[tb]
\leavevmode
\begin{center}$
\begin{array}{cc}
\scalebox{.5}{\includegraphics{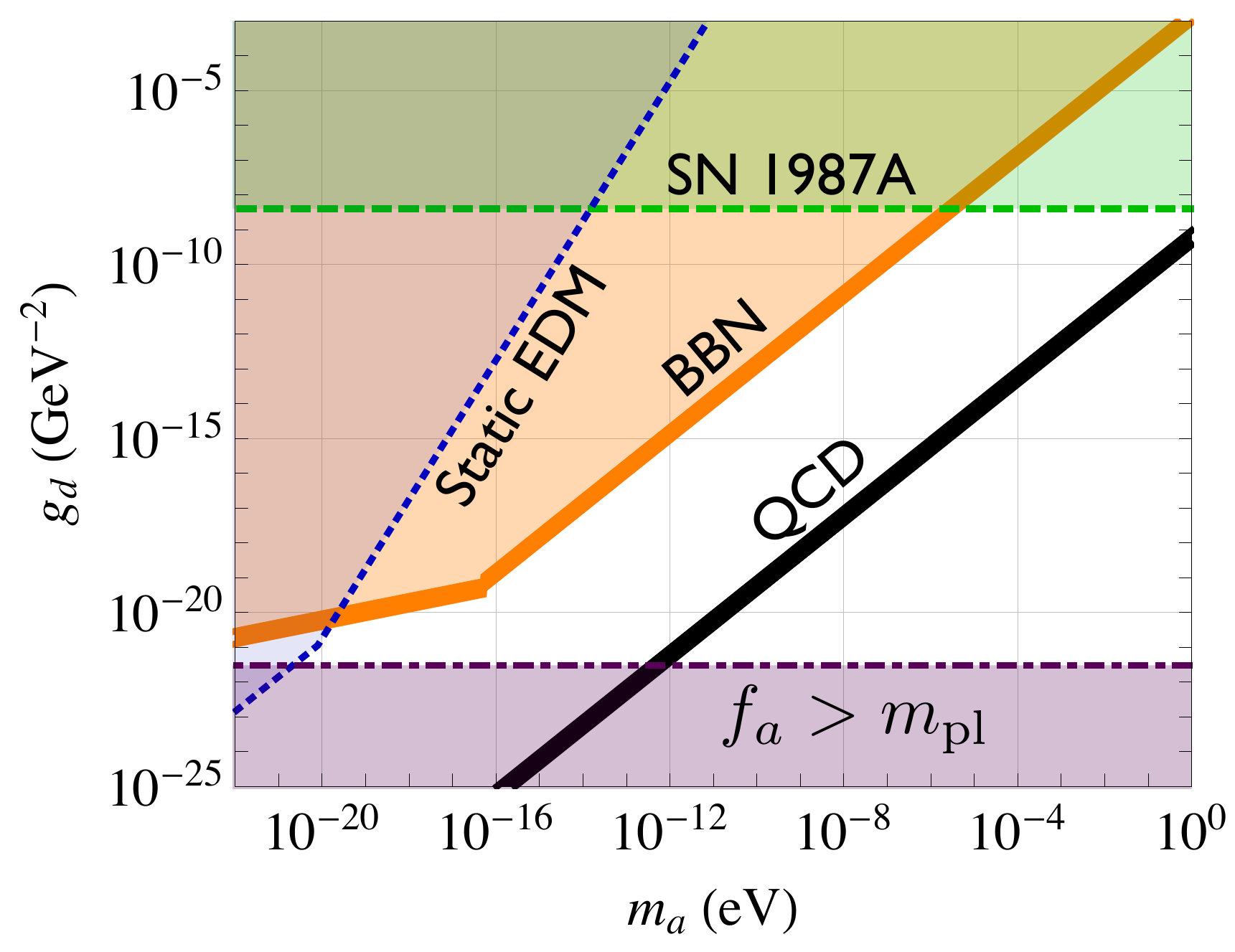}} & \scalebox{.5}{\includegraphics{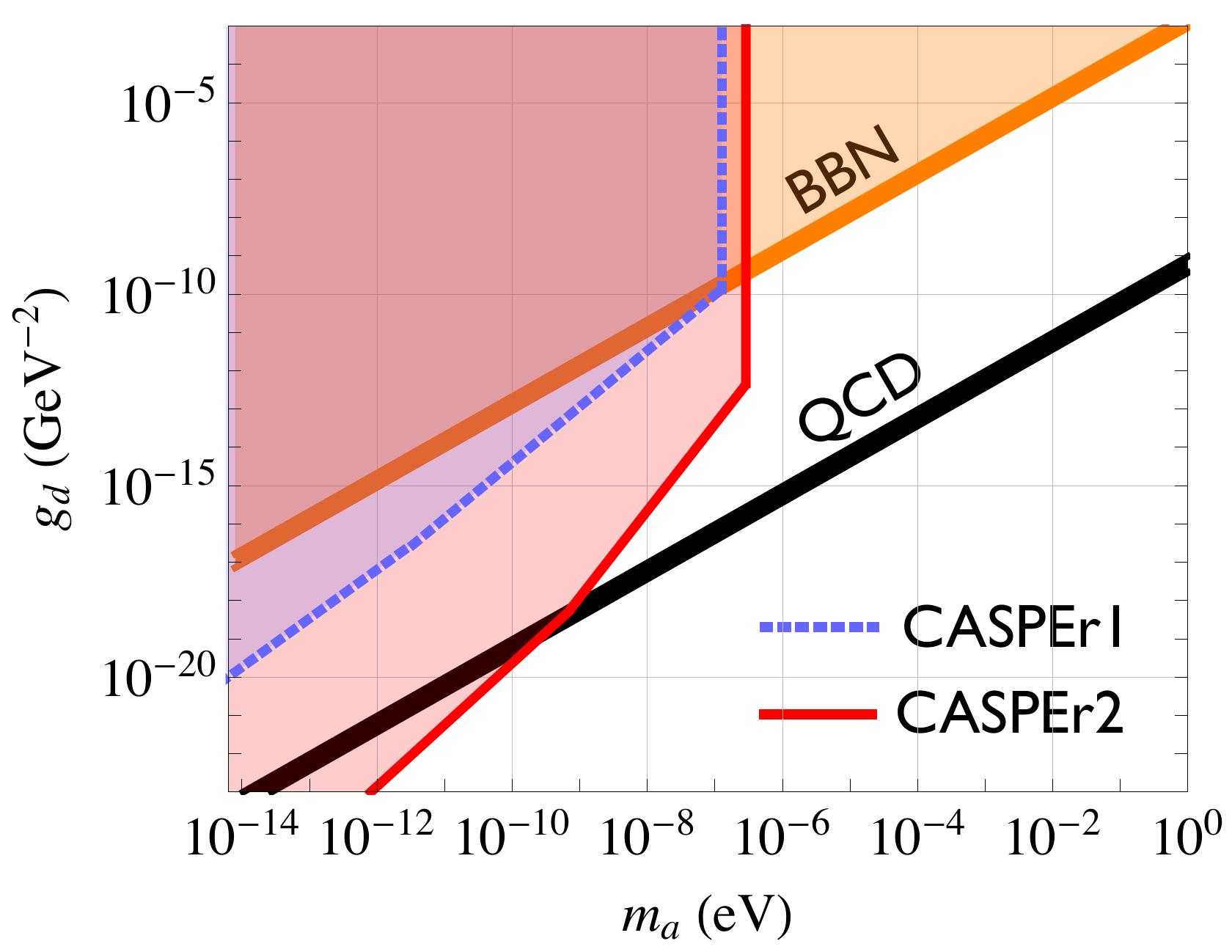}} \\
 \end{array}$
\end{center}
\vspace{-.50cm}
\caption{Left panel: BBN excluded region in the ($g_d$, $m_a$) plane is shown in orange. Other constraints include static EDM searches (blue shaded region, dashed blue boundary) and the bound from SN~1987A estimated conservatively in~\cite{Raffelt:2006cw,Graham:2013gfa} (green shaded region). The shaded purple region with dot-dashed boundary denotes $f_a > m_{\rm pl}$. Right panel: The future projected sensitivity of the oscillating EDM search of Refs.~\cite{Budker:2013hfa,Graham:2013gfa}.  CASPEr1 and CASPEr2 are the first and second generations of the experiments, respectively. The black line in both panels represents the QCD axion, $f_a\,m_a \approx\Lambda_{\rm QCD}^2$.}
\vspace{-0.15in}
\label{fig: final}
\end{figure*}  

A more accurate numerical method for calculating $Y_p$ as a function of $f_a \, m_a$ involves integrating the rate equation for neutrons and protons (see, for example,~\cite{Bernstein:1988ad,Weinberg:2008zzc}). 
The rates for neutron $\Leftrightarrow$ proton conversion are modified in the presence of axion DM because of the correction $\delta Q$ to the neutron-proton mass difference.  For \mbox{$m_a > 10^{-16}$ eV}, we solve the rate equation using the time-averaged rates, where the averaging is performed over the axion oscillation time $\sim m_a^{-1}$.  

Below the freeze-out temperature, it is also important to include the effect of the deuterons, because $^4$He production proceeds through reactions that involve deuteron production.  The deuteron fraction is highly suppressed until the temperature goes sufficiently below the deuteron binding energy $\epsilon_D$, a phenomenon known as the deuterium ``bottleneck."  A semi-analytic calculation shows that the deuterium bottleneck ends when $T \approx \epsilon_D / 26\sim0.1\text{ MeV}$, at which point nearly all of the remaining neutrons are rapidly bound into $^4$He.      
The dependence of $\epsilon_D$ on $\theta$ is not known, and this means that we do not know if axion DM delays or speeds up the end of the deuterium bottleneck.  
However, we expect this effect to be sub-leading.  The reason is that by the time the universe has cooled to temperatures of order $0.1$ MeV, $\theta_\text{eff}$ is only $\sim$ 4\% of its value at $T_F$.  In addition, the effect of free neutron decay on the $^4$He abundance is small.    

The results of the numerical calculation are as follows.  At large $f_a \, m_a$, we find $Y_p \approx 0.247 $, which is consistent both with the observed abundance and with the more precise numerical calculations~\cite{Beringer:1900zz}.  
We find that $\delta Y_p / Y_p \approx 10$\% when $f_a \, m_a \approx 1.3 \times 10^{-9}$ GeV$^2$, confirming the analytical estimate.     

The results above pertain to $m_a>10^{-16}$~eV. For $m_a \ll 10^{-16}$ eV, the axion field is approximately constant during BBN and does not redshift up with increasing temperature, due to the Hubble friction.\footnote{Of course, when $m_a$ is similar in size to the Hubble parameter at freeze-out ($m_a \sim 10^{-16}$ eV), the calculation is more complicated.  We save the details for future work.}  In this regime, the $^4$He BBN bound is approximately 
\es{boundLow}{
f_a\, m_a &\gsim \sqrt{2} \times {\big(1.3 \times 10^{-9} \, \, \text{GeV}^2 \big)} \left( \frac{1+z_m}{1+z_F} \right)^{3/2} \\
&\approx \big(1.8 \times 10^{-9} \, \, \text{GeV}^2 \big) \left( {m_a \over 10^{-16}~\text{eV}} \right)^{3/4}   \,,
}
where $z_F$ is the redshift at neutron freeze-out and $z_m$ is defined via $H(z_m)=m_a$.

We now discuss the implications for the axion DM-induced EDM experiments proposed in Refs.~\cite{Graham:2011qk,Budker:2013hfa,Graham:2013gfa}. 
The authors point out that the oscillating background field of axion DM induces an effective, oscillating neutron EDM, 
\es{dnExp}{
&d_n(t) = g_d \, a(t)=\frac{\sqrt{2\rho_{\rm DM}}}{g_d^{-1}\,m_a}\,\cos (m_at)\,,  \\
& g_d\approx { \big(2.4 \times 10^{-16}  \, \, \text{e}~\text{cm} \big) \over f_a}  \,.
}    
For the QCD axion, the amplitude of the oscillating EDM is \mbox{$d_n\sim10^{-34}$~e cm}, assuming a local DM density  of \mbox{$\rho_\text{DM}\approx0.3$~GeV/cm$^3$}.  The experiment proposed in~\cite{Budker:2013hfa,Graham:2013gfa} detects this small, oscillating nuclear EDM using NMR techniques, and the prospective sensitivity is shown in the right panel of Fig.~\ref{fig: final} by the regions above the blue dashed and red solid lines.  

Fig.~\ref{fig: final} shows the region (orange) of the $g_d$, $m_a$ parameter space that is excluded by the $^4$He abundance from BBN.  The width of the solid orange line takes into account the roughly $40$\% uncertainty in the expression for $g_d$ in~\eqref{dnExp}~\cite{Pospelov:1999ha}.  The solid black line shows the prediction for the QCD axion, which lies safely below the BBN bound.  Static EDM searches exclude the region to the left of the blue dashed line~\cite{Baker:2006ts,Harris:1999jx,Graham:2013gfa}, and a conservative bound from SN~1987A excludes the region above the green dashed line~\cite{Graham:2013gfa,Raffelt:2006cw}.  Model-dependent constraints also arise from the axion's coupling to $E \cdot B$ (not shown, see~\cite{Beringer:1900zz}).

Our BBN analysis neglects the temperature dependence of the axion mass. If such temperature  dependence is important, then in the parameter space defined by~\eqref{firstSpace} the axion mass may go negative at some time between now ($z=0$) and BBN ($z\sim10^{10}$). In this case, BBN would see a value of $\theta$ dependent on the extra PQ breaking dynamics, regardless of its value today. Thus $\theta$ would naturally be $O(1)$, strengthening the bound. Alternatively, if the QCD-induced contribution to the axion mass increases between today and $T_F$, the BBN bound is weakened.  Resolving this issue requires understanding the axion mass at temperatures significantly below $T_F$. 
 
To conclude, we showed that the production of $^4$He during BBN provides a novel constraint on the coupling of axion DM to QCD.  In particular, BBN excludes a large region of axion DM parameter space, with implications for current and future searches for axion DM-induced nuclear EDMs.  
Our bound is conservative, allowing for $10\%$ deviation in the predicted amount of $^4$He and ignoring deviations in the abundances of other light elements, such as deuterium. Moreover, we reviewed the fact that if an axion lives anywhere above the black line in Fig.~\ref{fig: final}, then the strong CP problem is reintroduced and made worse.

Axion DM that couples to QCD induces operators in the chiral Lagrangian that redshift up in the early Universe. For $m_a\,f_a\sim10^{-9}$~GeV$^2$, the perturbation parameter $a/f_a$ that controls these operators approaches order unity at the time of BBN, even though it is negligible today. As a result, much of this parameter space is excluded. 
It would be interesting to investigate other constraints on these operators that may arise from astrophysics.

\noindent
{\it We thank Nima Arkani-Hamed, Peter Graham, Samuel Lee, Aaron Pierce, Surjeet Rajendran, Juan Maldacena, Edward Witten, and Matias Zaldarriaga for useful discussions. KB was supported by the DOE grant de-sc0009988. BRS was supported by the NSF grant PHY-1314198. RTD was supported by the NSF grant PHY-0907744.}
\onecolumngrid
\vspace{1.0mm}
\twocolumngrid
\bibliographystyle{apsrev}
\bibliography{DMmod}

\end{document}